\documentclass{aastex}
\begin{document}
\title{The Saturation Limit of the Magnetorotational Instability}
\author{Ethan T. Vishniac}
\affil{Department of Physics and Astronomy, McMaster University, Hamilton ON L8S 4M1, Canada}
\email{ethan@mcmaster.ca}
\begin{abstract}
Simulations of the magnetorotational instability (MRI) in a homogeneous shearing box have shown that  the asymptotic strength of the magnetic field declines steeply with increasing resolution. Here I model the MRI driven dynamo as a large scale dynamo driven by the vertical magnetic helicity flux.  This growth is balanced by large scale mixing driven by a secondary instability. The saturated magnetic energy density depends almost linearly on the vertical height of the typical eddies.  The MRI can drive eddies with arbitrarily large vertical wavenumber, so the eddy thickness is either set by diffusive effects, by the magnetic tension of a large scale vertical field component, or by magnetic buoyancy effects.  In homogeneous, zero magnetic flux, simulations only the first effect applies and the saturated limit of the dynamo is determined by explicit or numerical diffusion.  The exact result depends on the numerical details, but is consistent with  previous work, including the claim that the saturated field energy  scales as the gas pressure to the one quarter power (which we interpret as an artifact of numerical dissipation).    The magnetic energy density in a homogeneous shearing box will tend to zero as the resolution of the simulation increases, but this has no consequences for the dynamo or for  angular momentum transport in real accretion disks.  The claim that the saturated state depends on the magnetic Prandtl number may also be an artifact of simulations in which microphysical transport coefficients set the MRI eddy thickness.  Finally, the efficiency of the MRI dynamo is a function of the ratio of the Alfv\'en velocity to the product of the pressure scale height and the local shear.  As this approaches unity from below the dynamo reaches maximum efficiency.  Farther from the disk midplane the Parker instability will dominate the local dynamics and the dynamo process.

 \end{abstract}
 \keywords{accretion, accretion disks,magnetic fields,
(magnetohydrodynamics:) MHD }
\section{Introduction}


For almost fifty years we have known that magnetic fields embedded in a differentially rotating flows with a negative radial gradient for the angular velocity are linearly unstable to radial perturbations \citep{V59,C60}.  This instability attracted only a modest level of interest among astrophysicists until the early 1990's when Balbus and Hawley (Balbus \& Hawley 1991; Hawley \& Balbus 1991;  for a review see Balbus \& Hawley 1998) pointed out that the magnetorotational instability (MRI) was uniquely suited to driving angular momentum transport in accretion disks.  It is a robust linear instability, in the sense that the maximum growth rate is close to the local shear rate, regardless of the strength of the magnetic field and it feeds directly off the energy released by the outward transport of angular momentum.  In the nonlinear limit, the angular momentum flux generated from this instability is proportional to the magnetic energy density.  This is, in turn, determined by the dynamo activity generated from the turbulence.  Determining how that dynamo operates, and the consequent mean value of $B^2$ in an accretion disk, then became a central issue in understanding ionized accretion disks.

Early work on this problem tended to view the dynamo as a side effect of the turbulence, with the magnetic field and the kinetic energy building up together and at similar scales.  This would naturally tend to saturate as the magnetic energy density approached the thermal pressure \citep{BH98,O03}.  Alternative models, inspired largely by phenomenological evidence from the study of cataclysmic variables \citep{MMH83}, looked for dynamo effects which would be sensitive to the geometry of the disk, i.e. the ratio of the disk height to radius.  Such models included dynamos driven by waves generated by tidal interactions at large radii \citep{VD92} and the incoherent dynamo \citep{VB97} in which the large scale field is built up by random contributions to the electromotive force from individual eddies.  However, subsequent three dimensional numerical simulations of the MRI have failed to show evidence for a dependency on box geometry.  More disturbingly, simulations tend to show evidence for saturation when the magnetic energy density was less than $10^{-2}$ times the ambient pressure \citep{BNST95,SHGB96}.  Given phenomenological arguments in favor of a dimensionless viscosity close to $0.1$ in dwarf novae \citep{BP81}  these results were somewhat disappointing.  On the other hand, simulations with vertical structure have tended to saturate at substantially  larger values \citep{MS00, HK01}.   This naturally raises the question of the role of magnetic buoyancy in the dynamo, both in terms of its effects on the MRI and the possibility that the Parker instability may contribute to the dynamo \citep{TP92, JL08}.

More recently there has been an attempt to understand the saturation value of the MRI driven dynamo by returning to the simplest possible simulations, shearing boxes with no vertical gravity and no imposed vertical field \citep{SITS04,FP07}.  These simulations show a steep dependence on numerical resolution, in the sense that at late times the magnetic energy density is inversely proportional to the number of grid elements.  The obvious conclusion is that in the limit of very high resolution, or more physically, in the limit of very low viscosity and resistivity (and absent some externally imposed vertical field), the MRI driven dynamo is  a negligible effect \citep[for a discussion of this possibility see][]{PCP07}.  

In this paper we reexamine the nature of the MRI dynamo.  We start from the assumption that it is a mean field dynamo effect, driven the magnetic helicity flux created by the MRI.  In \S 2.1 we derive the dynamo growth rate.  In \S 2.2 we find the dominant dissipative effect, which turns out to be driven by a secondary instability driven by vertical gradients in the large scale magnetic field and the angular momentum transport driven by the MRI.  In \S 2.3 we show how the typical MRI eddy thickness controls the strength of the magnetic field and discuss how this affects simulations and realistic accretion disks.  In particular, we note that magnetic buoyancy provides a minimal eddy thickness for MRI turbulence and provides a mechanism for maintaining an MRI dynamo in a zero mean flux accretion disk.  We briefly discuss the possible role of the Parker instability at large distances from the disk midplane.  Finally, in \S 3 we summarize our results and briefly comment on ways to test this model.  We comment, where appropriate, on the possible role of an imposed vertical field, but our treatment is aimed primarily at understanding how the disk dynamo functions when such environmental effects are negligible.

\section{Scaling Relations for the MRI Dynamo}

\subsection{The Growth Rate for the MRI dynamo}

We consider simplest example of the MRI dynamo, a disk with no entrained magnetic field.  In other words, the vertically averaged toroidal flux is zero and there is no imposed vertical field.  In the limit where the vertical wavelength is much smaller than the vertical scale heights of pressure, density or magnetic field, the linear dispersion relation has can be found in \citet{VD92}.  The presence of shear destroys the normal modes of the system, but they can be recovered as approximate solutions in certain limits.  A detailed analysis can be found in \citet{J07}.   For our purposes the essential points are: (1)  for $k_z>>k_\phi$ one recovers the usual MRI dispersion relation \citep{BH91} with the characteristic frequency expressed as ${\bf k}\cdot {\bf B}/\sqrt{4\pi\rho}$, (2) in that limit the growth rate does not depend on $k_z$,  (3) a locally unstable solution will grow for roughly $k_z/k_\phi$ e-foldings and (4) the characteristic radial wavenumber is of order $\Omega/V_A$ for all growth rates, where $\Omega(r)$ is the angular velocity (and also the local shear) and $V_A$ is the Alfv\'en speed.  The MRI dispersion relation predicts a growth rate $\sim \sqrt{3} k_\phi V_A$ for small $k_\phi$, a peak at some large fraction of $\Omega$ and a sharp cutoff at higher $k_\phi$.  The net result is that we expect MRI driven turbulence to be characterized by radial and azimuthal wavenumbers of order $\Omega/V_A$, much larger vertical wavenumbers, and a coherence time, $\tau$,  a few times $\Omega^{-1}$.  Numerical simulations are generally consistent with these expectations \citep{BNST95,SHGB96}, with eddies that are significantly stretched in the azimuthal direction, consistent with the idea that $\Omega\tau$ is larger than one by some modest factor.

Here we will adopt the hypothesis that the MRI drives a dynamo through the creation of a magnetic helicity flux.  The basic physics behind this kind of dynamo is set forth in \citet{vc01}\citep[see also][]{BS05,SB06}.  Here we will briefly summarize the concept and the key results.

Neglecting resistivity, the magnetic helicity, ${\bf A}\cdot{\bf B}$, is a conserved quantity.  Even when resistivity is not negligible the tendency of magnetic helicity to undergo an inverse cascade means that it is approximately conserved, a result that has major implications for laboratory experiments \citep{T86}.  If we apply a high pass filter we can rewrite the magnetic helicity conservation equation in the form
\begin{equation}
\partial_t h+{\bf \nabla}\cdot{\bf j}_h+2{\bf B}\cdot\langle {\bf v\times b}\rangle=-2\eta\langle{\bf j}\cdot{\bf b}\rangle,
\label{hel}
\end{equation}
where $h\equiv \langle{\bf a}\cdot{\bf b}\rangle$ is the magnetic helicity contained in eddy scale structures, $\eta$ is the ohmic resistivity, and ${\bf j}_h$ is the magnetic helicity flux contained in eddy scale structures.   The right hand side describes the effects of dissipation and is usually negligible in astrophysical contexts. The third term on the left hand side of this equation gives the transfer of magnetic helicity between scales.   This  expression can be used to constrain the electromotive force, $\langle{\bf v\times b}\rangle$.  The electromotive force contains a term proportional to the current helicity, $\langle{\bf j}\cdot{\bf b}\rangle$, which is $\approx k^2h$ in the Coulomb gauge.  The effect of this term is to transfer magnetic helicity from small scales to large at a rate which is $\sim (B^2/b^2)$ times the eddy turn over rate.  When this rate is faster than the dynamo growth rate we can neglect the $\partial_th$ term and take
\begin{equation}
\langle{\bf v\times b}\rangle_{\|}\approx {-2\over B}\nabla\cdot{\bf j}_h,
\end{equation}
where the subscript $\|$ denotes the component parallel to the large scale magnetic field.
For the MRI we have $\langle v^2\rangle\sim \langle b^2\rangle\sim B^2$ so we are always in this limit.
When ${\bf j}_h$ is zero, or negligibly small, we get the phenomenon of "$\alpha$ quenching" \citep{gd94}.

In the presence of differential rotation, ${\bf j}_h$ is not zero.  On dimensional grounds it has to be proportional to the square of the magnetic field times a transport coefficient.  Neglecting constants of order unity, quasilinear  estimates from \citet{vc01} give
\begin{equation}
{\bf j}_h\sim D_z B^2 \Omega\tau\hat z.
\end{equation}
If we follow convention and write the electromotive force as $\alpha{\bf B}$ then this is equivalent to
\begin{equation}
\alpha\sim {D_z\over L_B}\Omega\tau,
\end{equation}
where $L_B$ is the vertical scale for the large scale magnetic field.
This implies a growth rate of
\begin{equation}
\Gamma\sim\sqrt{{\alpha\over L_B}\Omega}\sim {k_{\|}\over k} \Omega\tau {v_{\rm turb}\over L_B},
\end{equation}
where $L_B$ is the vertical scale length for the large scale magnetic field.  For the MRI we have
$\Omega\tau\sim 1$, $v_{\rm turb}\sim B$, and $k_{\|}B\sim \Omega$.  We get a  dynamo growth rate of
\begin{equation}
\Gamma\sim {\Omega\over k L_B}.
\label{growth}
\end{equation}
For typical eddies we have  $k\sim k_z\gg \Omega/V_A$.  However, since the magnetic helicity flux goes as $k^{-2}$, it is possible, even likely, that it will be dominated by modes with $k_z$ close to $\Omega/V_A$, depending on the power spectrum of the MRI driven turbulence.

A particular solution for this kind of dynamo was discussed in \citet{vc01}, i.e. a dynamo in a periodic shearing box.  This is appropriate for the MRI driven dynamo simulations we are discussing here.  Its applicability to real disks, i.e. non-periodic systems is less clear.  This solution is characterized by a constant vertical magnetic helicity flux, oscillating between ${\bf j}_h$ and large scale field components.  In a real disk a necessary part of the dynamo process is the ejection of magnetic helicity from the top and bottom of the disk.  However, in contrast to earlier suggestions \citep[see, for example][]{BF01}, the ejected magnetic helicity can be contained in large scale structures or eddy scale structures and the net magnetic helicity of the system does not need to change.  Since ejecting magnetic helicity bound up in large scale structures requires much less energy it is likely to be the preferred mode of ejection. 

The implication of this work is that efficient dynamo action is possible even in a periodic box provided that the dynamo process is dominated by the magnetic helicity flux.   There is some work exploring the effect of boundary conditions.  We note in particular the work of \citet{TCB08} and \citet{KKB08}.  In both cases they found that periodic boundary conditions were significantly less favorable for the development of a large scale field than periodic boundary conditions.  However, both simulations were in the limit of very weak large scale field where the magnetic helicity driven dynamo is unlikely to operate.  In contrast, MRI simulations are always in the strong field limit.  Finally, we note that recent work exploring the helicity driven dynamo in a periodic box with externally driven small scale turbulence has found large parts of parameter space with strong dynamo growth and field saturation with $\langle B^2\rangle>\langle b^2\rangle$ \citep{SV09}.

\subsection{Dissipation of the large scale field}
The usual estimate for the turbulent dissipation coefficient in arbitrary direction $\hat i$ is
\begin{equation}
{\cal D}_i\sim \langle v_i^2\rangle \tau\sim {1\over k_i^2\tau},
\end{equation}
where $k_i^{-1}$ is the correlation length in the $\hat i$ direction.  In the quasilinear approximation we 
equate this wavenumber with the wavenumber of the dominant eddies driven by the local instability.
For the MRI this implies a vertical diffusion coefficient which vanishes in the limit $k_z\rightarrow\infty$.
In fact, the resulting turbulent diffusion rate,
\begin{equation}
\tau_{\rm diss}^{-1}\sim \tau^{-1}{1\over (k_z L_B)^2},
\end{equation}
vanishes faster than the dynamo growth rate given in equation (\ref{growth}).  The implication is that the vertical turbulent diffusion of the large scale magnetic field, driven by the MRI,  is a negligible effect for the high resolution shearing box simulations.

If turbulent diffusion driven by the MRI is ineffective in this case then we need to consider other, normally subdominant effects.  Given a scalar pressure and a homogeneous environment there are none.  However, even in the simplified environment provided by a homogeneous shearing box simulation of the MRI both these assumptions are violated.  First, the stress tensor includes an anisotropic component of $ \rho v_iv_j-b_ib_j (4\pi)^{-1}$.  If we average this over length and time scales typical of the MRI  we are left with  significant terms in the $\hat r\hat r$, $\hat\phi\hat\phi$, and $\hat r\hat\phi$ directions, all positive contributions of order $\rho V_A^2$ (see, for example, Stone et al. 1996). Second, magnetic energy stored in large vertical wavelength modes provides a natural way to break the homogeneity of the environment.  Finally, we can point to the simulations of \citet{JY08} which show the appearance of large scale, axisymmetric, slowly evolving structures in MRI turbulence in stratified and unstratified shearing box simulations.  Evidently there is  the potential for large scale vertical mixing.

To gain some understanding of how these structures arise we can look for linear modes with slow growth rates.  Shearing imposes the condition 
\begin{equation}
k_r\gg k_\phi {\Omega\over\omega},
\label{shear}
\end{equation}
where $\omega$ is the magnitude of the complex frequency of our hypothetical large scale secondary instability.   This suggests that we should concentrate on axisymmetric modes, or at least modes for which $k_\phi$ is negligible. Treating the MRI  only as a homogeneous process that generates extra terms in the stress tensor we see that these terms  depend on the square of the  magnetic field.  Assuming that the dynamo process is slow and that the magnetic field evolves only by the advection of the large scale field we can generate radial gradients in the extra components of the stress tensor by moving fluid elements vertically.  If $P_{ij}\equiv C_{ij}\rho V_A^2$ then our equations of motion are:
\begin{equation}
\partial_t v_\phi +{\Omega\over2}v_r =-ik_r C_{r\phi} 2 V_A \delta V_A,
\end{equation}
\begin{equation}
(\partial_t +D_{rr} k_r^2) \delta V_A=-v_z\partial_z V_A,
\end{equation}
\begin{equation}
\nabla\cdot\vec v=0,
\end{equation}
\begin{equation}
\partial_tv_z={-1\over\rho}\partial_z \delta P,
\end{equation}
and
\begin{equation}
\partial_t v_r -2\Omega v_\phi=-ik_r(\delta P +C_{rr }2V_A\delta V_A).
\end{equation}
Here  $\delta P$ is the perturbation to the scalar pressure, $\delta V_A$ is the perturbation to the Alfv\'en velocity, the magnetic field is assumed to be entirely in the $\hat \phi$ direction, and $D_{rr}$ is the radial turbulent diffusion coefficient.  Given that we expect $\omega<<\Omega$ we can drop the $\partial_t v_\phi$ term in the first equation and combine the remaining equations to obtain
 \begin{equation}
(\partial_t+D_{rr}k_r^2)\partial_z v_z \approx {2k_r^2 C_{r\phi}\over\Omega}v_z \partial_z V_A^2.
\label{ripple1}
\end{equation}
We also note that 
\begin{equation}
D_{rr}\approx C_{rr}V_A^2\tau,
\label{rdiss}
\end{equation}
where $\tau$ is the characteristic correlation time of the MRI turbulence.  If we ignore $D_{rr}$ then
we can find periodic traveling wave solutions to this equation with wavelengths less than the wavelength of the magnetic field.  Multiplying both sides of equation (\ref{ripple1}) by $\partial_z v_z$ we see that if $v_z\propto V_A$ then when dissipation is small $\partial_z v_z$ increases monotonically.  This is not the same as proving that we have an instability, but it indicates that  taking $v_z\propto V_A$ gives the best chance of an instability and that the traveling wave solutions are irrelevant.    In order to gain further insight we take $V_A\propto V_0\cos(\kappa z)$, and define two characteristic rates:
\begin{equation}
\gamma_r\equiv {k_r^2V_0^2C_{r\phi}\over\Omega},
\end{equation}
and
\begin{equation}
\gamma_d\equiv k_r^2V_0^2C_{rr}\tau.
\end{equation}

The ratio of $\gamma_r$ to $\gamma_d$ 
can only be determined by appeal to numerical simulations, and $\tau$ is difficult to estimate from the published simulations.  However, it should be equivalent to measuring the turbulent magnetic Prandtl number, or more properly the Schmidt number.  This can be measured in simulations and the results have been reported in several simulations \citep{CSP05, JK05, TWBY06, FP06, JKM06}.  The measurement varies from about 11 \citep{CSP05} to values of 0.8-0.9 \citep{JK05}.  \citet{JKM06} pointed out that the value increases sharply with the  applied vertical field and has a different dependence for the radial and vertical Schmidt numbers.  For our purposes we need the former at small vertical field strengths, which brings us back to the values quoted in \citet{JK05}.     

If we take $v_z(t=0)=\cos(\kappa z)$ then we can expand the solution at all times as
\begin{equation}
v_z=\sum_{n=1}^\infty a_n(t)\cos[(2n-1)\kappa z].
\end{equation}
In Fig. 1 we plot the $a_1(t)$ for $\gamma_d=0,0.1\gamma_r,$ and $ \gamma_r$.  The second value is a reasonably close approximation to the results of Carballido et al. whereas the last value is consistent with the work of \citet{JK05}.    In all three  cases $a_1$ increases monotonically, and almost exponentially.  The case with no dissipation does not look like a pure mode, but the other two cases do.  In all cases the higher order coefficients are positive, although smaller, and show similar behavior.  These are not  pure exponential modes, but the difference does not seem significant.  It is interesting to note that a small amount of radial mixing actually enhances the growth of this instability.   
\begin{figure}
\plotone{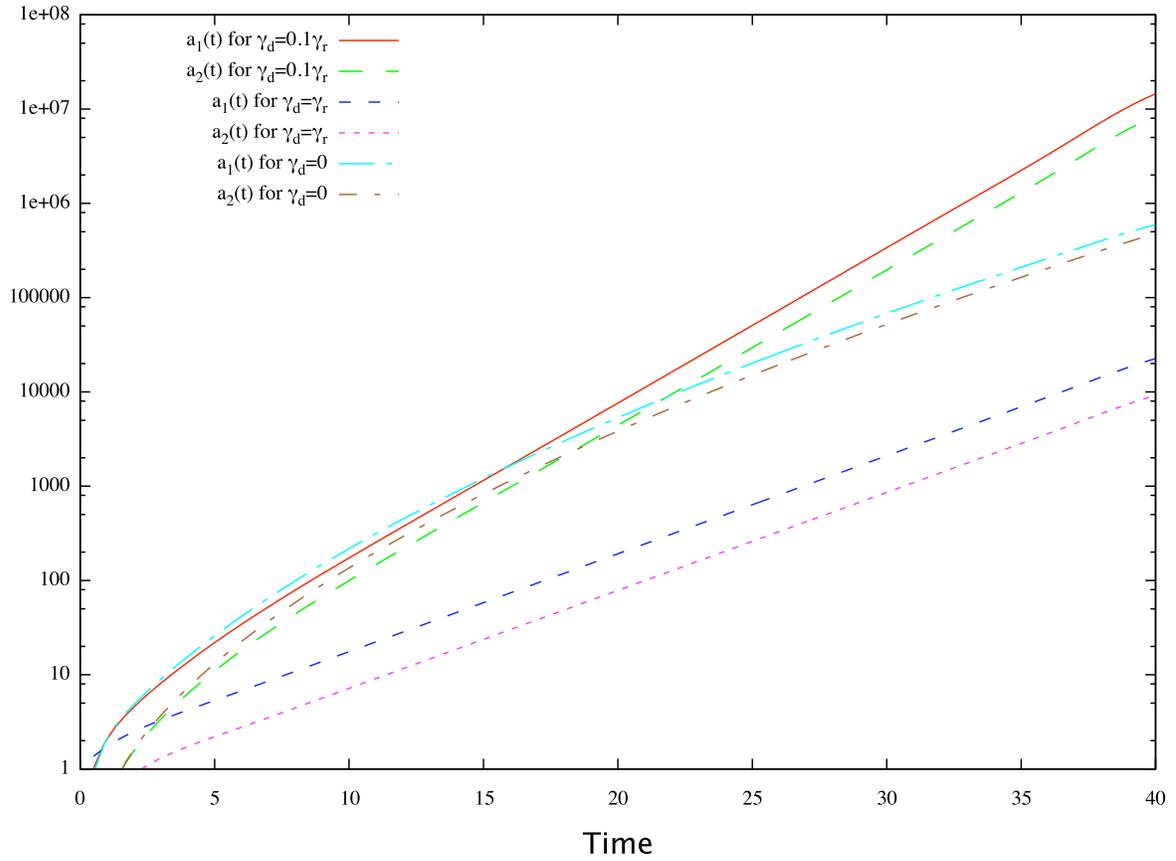}
\caption{The two leading Fourier coefficients for the rippling modes as a function of time for varying levels of dissipation.}
\end{figure}

This linear model includes the assumption that the dynamo process is very slow.  Since we are looking for a mixing process that will balance the dynamo, this is clearly unrealistic.  Rather by taking this limit we
are aiming for an estimate of the dissipation time scale.  The actual dynamics of the magnetic field , with
the dynamo and the secondary instability in statistical equilibrium, will require a more inclusive treatment,
which is beyond the scope of this paper.

What do we expect in the nonlinear limit?  Taking $k_rL_B\sim1$, 
 we should get circulation over vertical scales of order $L_B$ on times of order $L_B^2\Omega/V_A^2$.  This gives rise to an effective turbulent transport coefficient of roughly $V_A^2\Omega^{-1}$.  This is just what we would have obtained if the MRI gave rise to approximately isotropic turbulence.  The corresponding rms vertical velocity is just
\begin{equation}
V_{\rm ripple}\sim L_B\gamma_r\sim {V_A^2\over L_B\Omega},
\end{equation}
which will not affect local measurements of the rms vertical velocity provided
\begin{equation}
V_A^2\le {L_B\over k_z}\Omega^2.
\end{equation}
We will show below that this is the expected saturated limit for the magnetic field driven by the MRI.  Since these modes are largely two dimensional, we can expect that they will show signs of inverse cascade, building coherent flows on the largest scales available.  Furthermore, the inertial term in the $\hat z$ component of the force equation will produce a contribution to $v_z$ which is proportional to $\partial_z v_z^2$.  Since the growing mode has a large component of $v_z$  proportional to $V_A$ this implies a nonlinear contribution to $v_z$ which is proportional to $\partial_z V_A^2$.
When we substitute this into the RHS of equation (\ref{ripple1}) we get a term which is independent of $z$.  This cannot be balanced by the LHS of the same equation.  Instead, we need to appeal to the neglected term, $\partial_tv_\phi$.  In other words, we expect this instability to result in the appearance of annular bands of fast and slow rotation, supported by radial pressure gradients.  These variations in orbital velocity will be comparable to the vertical and radial motions.

Finally, we note that the slow motions driven by this rippling effect are not necessarily axisymmetric.  Applying the condition in equation (\ref{shear}) we have
\begin{equation}
k_\phi < {1\over L_B}\left({V_A\over L_B\Omega}\right)^2.
\end{equation}
Taking $L_B$ to be something like the disk thickness, so that the RHS of this equation is roughly $\alpha_{SS}\over H$, we see that these modes will be axisymmetric unless the disk radius is greater than $H\alpha_{SS}^{-1}$.  In the homogeneous shearing box simulations these modes will have effectively no $\phi$ dependence.

In terms of the evolution equation for the large scale field, we can include this effect by adding a term $-D_{\rm ripple}{\bf J}\sim -V_A^2\Omega^{-1}{\bf J}$ to the electromotive force.  This is not quite right, because this creates an extra component of the electromotive force parallel to the large scale magnetic field.  This, in turn, violates magnetic helicity conservation.  In order to restrict the injection of magnetic helicity into the large scale field to the inverse cascade from the small scale field we need to subtract out the part parallel to ${\bf B}$.  In other words, the correction to the electromotive force is
\begin{equation}
\Delta{\cal E}_{\rm ripple}\sim {1\over 4\pi\rho\Omega}\left[B^2{\bf I}-{\bf BB}\right]\cdot{\bf J}={1\over4\pi\rho\Omega}\left[({\bf J}\times{\bf B})\times{\bf B}\right].
\label{diff}
\end{equation}
We would get the same result if we simply  appealed to the force exerted by the large scale field and assumed that it was coupled to the fluid on an eddy turnover time (a heuristic model suggested by \citet{BS00}).  However, despite the apparent agreement between the two approaches there are significant physical differences between them.  In our approach the local shear $\Omega$ enters through the coriolis force.  In the ambipolar approximation to large scale field dynamics the quantity $\Omega^{-1}$ enters as the eddy turn over time.  For the MRI these give roughly the same time scale, but this is not true in general.  In addition, using the ambipolar approximation for the large scale field evolution in this way implies that we are including the pressure force exerted by the large scale field, despite the fact that the scalar pressure is almost unperturbed everywhere for $k_z$ very large.  In our approach this "pressure" term is actually a description of the effects of a slow large scale instability and the "tension" term the result of the back reaction from small scale magnetic field structures arising from magnetic helicity conservation.  It would appear that the resemblance between the two approaches is a coincidence, rooted in the properties of the MRI and magnetic helicity conservation.

For an idealized $\alpha-\Omega$ dynamo, with $B_z\approx 0$, $|B_\phi|\gg B_r$, and no radial or azimuthal derivatives, we can rewrite the RHS of equation (\ref{diff}) as
\begin{equation}
 \Delta{\cal E}_{\rm ripple}\approx [B_\phi^2\partial_z B_\phi\hat r -B_r^2\partial_zB_r\hat\phi-B_rB_\phi
\partial_zB_\phi\hat\phi]\tau.
\label{drift1}
\end{equation}
At first glance this looks like the turbulent dissipation of $B_r$ is heavily suppressed, by a factor of 
$\sim (B_r/B_\phi)^2$, but when $B_r$ and $B_\phi$ are strongly correlated, as expected in a shearing system, we can approximate $B_r\partial_zB_\phi\approx B_\phi\partial_zB_r$ so that equation
(\ref{drift1}) becomes
\begin{equation}
\Delta{\cal E}_{\rm ripple}\approx [B_\phi^2\partial_z B_\phi\hat r -B^2\partial_zB_r\hat\phi]\tau.
\label{drift2}
\end{equation}
Then the  dissipation rates for both components of the large scale magnetic field are 
\begin{equation}
\tau_{\rm diss}^{-1}\sim {V_A^2\tau\over L_B^2}
\end{equation}

Balancing this with the dynamo growth rate given in equation (\ref{growth}), and using $\Omega\tau\sim1$, we predict a saturated magnetic field energy density of
\begin{equation}
B^2\sim 4\pi \rho {L_B\over k_z}\Omega^2.
\label{sat1}
\end{equation}
This is not quite the end, because the magnetic helicity flux becomes sharply less efficient at higher vertical wavenumbers, so it could easily be dominated by the smallest $|k_z|$ modes.  As a first approximation, we will assume that the unstable modes of the MRI are all equally excited.  This is physically reasonable, and perhaps more important, it is consistent with the simulations of \citet{FP07} (see, in particular, their Fig. 6).  In this case the magnetic helicity flux will be dominated by the largest vertical wavelength modes ($k_z\sim \Omega/V_A$) but these will contain only a fraction $\sim k_z/k_c$ of the total energy, where $k_c$ is the small scale cutoff to the vertical wavenumber.  This implies an effective
vertical wavenumber
\begin{equation}
k_{\rm eff}\sim \left({k_c\Omega\over V_A}\right)^{1/2},
\label{keff}
\end{equation}
and a saturated state with 
\begin{equation}
B^2\sim 4\pi \rho L_B^{4/3} k_c^{-2/3}\Omega^2,
\label{sat}
\end{equation}
which we expect to apply in  the asymptotic limit where $k_cV_A/\Omega\gg1$.

Equation (\ref{sat}) is the main result of this paper, but it is physically incomplete.  The maximum vertical wavelength, $k_c$, driven directly by the MRI is not a fundamental parameter.  It is determined by effects other than the MRI itself.  In real accretion disks $k_c$ may be set by the magnetic tension associated with some large scale vertical field.  In stratified disks magnetic buoyancy will couple fluid motions at different heights and provide a lower limit to the thickness of MRI eddies.  Finally, viscosity and resistivity, either as real physical effects or as the consequence of limitations in numerical calculations, will impose a maximum vertical wavenumber.  In the next section we will discuss how each of these effects determine the strength of large scale magnetic fields, first in numerical simulations, then in real accretion disks.

\subsection{The vertical wavenumber in homogeneous simulations and in accretion disks}

In homogeneous numerical simulations, where there is no vertical gravity, the only limit on the vertical wavenumber comes from the "microphysics" of the simulation, i.e. the numerical effects which produce an effective small scale dissipation.  In the limit of infinite resolution, we expect $k_c\rightarrow\infty$ and $V_A^2\rightarrow 0$.  In order to understand simulations with finite resolution we need to consider the scaling laws that govern $B^2$ in the presence of numerical diffusion.  This requires appealing to the details of the specific numerical codes used to find the saturated state of the MRI in a homogeneous shearing background.  We will examine two recent attacks on this problem: \citet{SITS04}  and \citet{FP07}.  We start with the latter because they give a much more detailed summary of their results.  For their study Papaloizou and Fromang used the ZEUS code.   They reported the effect of changing resolution on the distribution of numerical dissipation in Fourier space.  While the plot of numerical dissipation versus wavenumber is a bit noisy, with a broad peak, they also included fits based on an equivalent physical dissipation.  These fits, presented in their figures   9, 10, and 11, show a peak around $k=\sim 25, \sim 50, \sim100$ for vertical grid spacings of 64, 128, and 256 respectively.  This allows us to avoid modeling the numerical dissipation and simply note that the critical value of $k_z$ scales almost inversely with grid spacing $\Delta$ in ZEUS.  They also plot the reduced power spectra for the kinetic and magnetic energies as a function of  $k_z$ for all three resolutions in figure 6.  This shows a relatively flat spectrum, in terms of the amplitude per vertical wavenumber, for all three resolutions, although the spectrum is flat for a significant range in $k_z$ only at the highest resolution.  These results confirm the notion that all the allowed modes in $k_z$ share equally in the MRI, and support the idea that we should weight the longest wavelength modes by their share of phase space, i.e. the factor $\Omega/(k_c V_A)$, when estimating the magnetic helicity flux.  On the other hand,   equation (\ref{sat}) predicts from this that the magnetic energy density and angular momentum flux should then scale as $\Delta^{2/3}$.  Fromang and Papaloizou quote a slightly steeper dependence, consistent with $B^2\propto \Delta$ which suggests that the magnetic helicity flux is dominated by the highest wavenumber modes.  If true, this would be odd.  It is more likely that once we consider the gradual flattening of the spectrum as the resolution is increased, the dependence on vertical wavenumber  is consistent with our prediction.  This issue could be settled by calculating the contribution to the magnetic helicity flux from different modes in a shearing box simulation.

Sano et al. used a second-order Godunov-type scheme of their own devising.  They did not report power spectra, or an effective vertical wavenumber cutoff.  However, we can still gain some insight from their results.  In their code the time step scales with the grid spacing divided by the magnetosonic velocity, which in this case is effectively the sound speed.  It can be problematic to interpret numerical dissipation in terms of an effective transport coefficient, but for this code the most reasonable approximation is to use $D=C_0 \Delta c_s$ in place of viscosity.  Here $C_0$ is an effective scaling for the Courant condition which decreases as the Courant condition becomes more stringent.  The code should be able to handle MRI driven eddies as long as $k_z^2D$ is less than the MRI growth rate, or
\begin{equation}
k_z\le \left({\Omega\over c_s \Delta}\right)^{1/2} C_0^{-1/2}.
\end{equation}
In a simulaton with a fixed average density this implies $k_c\propto P^{-1/4}$.  Sano et al. reported a scaling of   $V_A^2\propto P^{1/4}$, which implies that in their simulations as well the saturated magnetic energy density scales as $k_c^{-1}$ rather than $k_c^{-2/3}$.
We note that in a conference paper \citet{S05} reported that the magnetic energy and stresses scaled linearly in the box size and as the square root of the grid cell dimension.   This is expected from  our estimate for $k_c$ in their work and the use of equation (\ref{sat1}) rather than (\ref{sat}).  As before, this result is broadly consistent with our theoretical expectations, but leaves unclear whether the deviation from equation (\ref{sat}) is because $(k_cV_A/\Omega)^2$ is not sufficiently large in these simulations, or because the magnetic helicity flux is dominated by the highest $k_z$ modes.

There is one other issue that should be raised regarding homogeneous shearing simulations, the role of the magnetic Prandtl number.  When this is large, i.e. viscosity much greater than resistivity, then $k_c$ is determined by the viscosity and resistivity plays no role in our model.  We note that simulations \citep{FPLH07} show that the MRI driven dynamo operates at maximal efficiency in this regime.  On the other hand, when the magnetic Prandtl number is small, or of order unity, then the RHS of equation (\ref{hel}) is no longer negligible, magnetic helicity is no longer a robustly conserved quantity, and the efficiency of the dynamo will be reduced (as seen in \citet{FPLH07}).  Exactly how much it is reduced, and whether or not the dynamo remains viable requires us to consider the degree to which magnetic helicity will tend to cascade to higher wavenumbers instead of being transferred to very large scales.  This is beyond the scope of this paper.  We simply note that this effect will be large in any simulation in which resistivity is not much smaller than viscosity and in which $k_c$ is set by these constants.

In ionized accretion disks, where the MRI is likely to be the dominant process for angular momentum transport, the resistivity and viscosity are very small and the microphysical limit on $k_z$ is so large as to be uninteresting.  Consequently we can assume that we are in the asymptotic regime, and use equation (\ref{sat}).   There are two macrophysical effects which give much smaller limits on the vertical wavenumber.  First, if there is a large scale vertical field, possibly advected in from large radii, then $k_z$ is limited by the tension of the vertical field, i.e.
\begin{equation}
k_c\sim  {\Omega(4\pi\rho)^{1/2}\over B_z},
\label{vcrit}
\end{equation}
so that
\begin{equation}
B^2\sim \left({B_z^2L_B\over4\pi\rho\Omega^2}\right)^{1/3} r\pi \rho L_B \Omega^2.
\label{vert}
\end{equation}

In real disks there will be some vertical field as a consequence of the existence of large scale magnetic domains.  Since $\nabla\cdot{\bf B}$ we have
\begin{equation}
B_z\sim {L_B\over L_R} B_r,
\end{equation}
where $L_R$ is the radial magnetic domain size.  This has frequently been used to argue that $B_z\sim (H/R) B_r$ in accretion disks.  However, it is unreasonable to suppose that the typical magnetic domain in an accretion disk is as big the whole disk.  From equation (\ref{sat}) we can reasonably expect the dynamo to favor magnetic domains as thick as the disk, but as long as $L_R>L_B$ all radial domain sizes are equally likely.  Simple phase space considerations then suggest that $L_R$ should typically be only slightly larger than $L_B$.  Moreover, if a strong vertical field component can enhance the local dynamo action, this limit is strongly favored.  All of which leads to the expectation that typically $B_z$ is of order $B_r$ or roughly $B_\phi/(L_B k_{\rm eff})$.  Combining equations (\ref{keff}), (\ref{sat}), (\ref{vcrit}), and (\ref{vert}) we find that assuming that internally generated value of $B_z$ determines $k_c$ results in a trivial equality.  Any constraint imposed by $B_z$ is scale free, i.e. independent of the parameters of the problem.  Since we see no evidence for such a constraint in the numerical simulations, it is reasonable to conclude that it never comes into play.  This is not to say that it is completely irrelevant.  We will see that in a real disk the dynamo will operate with an efficiency which depends on the distance to the midplane.  The vertical component of the large scale magnetic field constitutes a vertically nonlocal effect.  An efficient dynamo operating a density scale height or more away from the midplane can boost the efficiency of the dynamo at the midplane of the disk.

Alternatively, we can consider the effects of vertical structure on the individual eddies.  The MRI produces fluctuations in the local magnetic energy density of order unity, operating on time scales which are much longer than the sound travel time across an eddy.  Consequently the gas pressure will vary by an amount $\delta P\sim B^2$ and there will be fractional variations in the density of order $B^2/P$.  This will give rise to differential buoyant accelerations within an individual eddy of order $(B^2/P)z\Omega^2$.  Since these accelerations will act coherently for an eddy turn over time, $\sim \Omega^{-1}$, this implies a minimal eddy thickness $\Delta z$ of order $(B^2/P)z$. or
\begin{equation}
k_c\sim {c_s^2\over V_A^2}{1\over z}\sim {L_P\Omega^2\over V_A^2},
\label{bkeff1}
\end{equation}
where $L_P$ is the pressure scale height.  Using equation (\ref{sat}) for the magnetic field strength gives
\begin{equation}
k_cL_B\sim \left({L_P\over L_B}\right)^3.
\end{equation}
This will be enormous near the disk midplane, but will decrease rapidly as we move away from it.  In terms of numerical simulations, this suggests that a background pressure scale height much larger than the size of the computational box will still be sufficient to remove the dependence on numerical resolution.  For example, the highest resolution simulations of Fromang and Papaloizou had $k_cL_B\sim 100$, so a background pressure scale height $\sim 4$ times the box size or less would have been sufficient to compete with numerical resolution in setting $k_c$.

Equation (\ref{bkeff1})  will be consistent with our assumption that $k\sim k_z$ if
\begin{equation}
{k_cV_A\over\Omega}\sim {L_P\Omega\over V_A}\gg 1.
\end{equation}
As $V_A$ approaches this limit from below the MRI eddies will become increasingly isotropic and the MRI dynamo will operate with increasing efficiency.  Using this same expression for $k_c$, the growth rate for the MRI dynamo is
\begin{equation}
\Gamma\sim {\Omega\over k_{\rm eff} L_B}\sim {V_A\over L_B}\left({V_A\over L_P\Omega}\right)^{1/2}.
\end{equation}
The saturation level of the magnetic field is just
\begin{equation}
B^2\sim (4\pi\rho) \left({L_B\over L_P}\right)^4 L_P^2\Omega^2.
\end{equation}

This last result shouldn't be taken too literally.  It is based on a local balance between dissipation and the dynamo.  The actual saturated state in a disk requires a global calculation, and must include the variation in the background density and temperature.  The sense of this result is that the dynamo process will be much less efficient near the midplane, and will also depend on the diffusion of toroidal and vertical magnetic flux generated at larger scale heights.   In the end, we expect that   the magnetic energy will saturate at some modest fraction of the vertically averaged gas pressure, and that a more detailed solution requires a realistic consideration of the vertical structure of the disk.   We note that an externally imposed vertical field will have an effect when $V_{A,z}\rightarrow V_A^2/(L_P\Omega)$,.  This means that it will have an impact first near the midplane.

We have also assumed that the magnetic pressure is smaller than the gas pressure, which is a necessary condition for the MRI \citep{KO00}.  We note that this ratio can be rewritten as 
\begin{equation}
{V_A^2\over c_s^2}={V_A^2\over L_P z\Omega^2}=\left({V_A\over L_P\Omega}\right)^2 {L_P\over z}.
\end{equation}
This constraint is more important than $V_A\ll L_P\Omega$ only near the disk midplane.  Since the MRI driven dynamo is least efficient near the disk midplane, this suggests that this constraint is not important. 

There is one other point worth mentioning in this context.  We have assumed that the MRI is the only instability operating in the disk.   However, in a stratified disk we  also expect the appearance of the Parker instability.  The linear theory \citep{S74, FT95} gives us a typical growth rate of $V_A/L_P$ for azimuthal wavelengths of at least $L_P$.  Consequently  equation (\ref{shear})  implies a radial wavelength comparable to $V_A/\Omega$.  For $V_A\ll L_P\Omega$ this leads to the expectation that the MRI will suppress the Parker instability through radial mixing \citep{VD92}, consistent with the numerical results of \citet{SHGB96} who first looked for signs of the Parker instability in a simulation of the MRI in a stratified disk.  On the other hand, when $V_A\gg L_P\Omega$ the Parker instability should be unaffected by the shearing of the disk.  The implication is that the parameter $V_A/(\Omega L_P)$ defines a continuum marked, at small values, by an MRI driven dynamo whose efficiency increases as $V_A\rightarrow L_P\Omega$, and at large values by a Parker instability driven dynamo.

\section{Conclusions}

In this paper we have proposed a model for the dynamo driven by the magnetorotational instability.  It implies that the saturation strength of the magnetic field, and the attendant Maxwell stresses, scale as $L_B^{4/3}k_c^{-2/3}\Omega^2$, i.e. slightly more steeply than linearly with  box height and slightly less so with the MRI eddy height.  The available simulations suggest that the dependence on each is linear.  We see this in the scaling with magnetic domain size in the work of Sano et al.  The scaling with $k_c$ is seen most clearly in  the work of \citet{FP07}.  We tentatively suggest that the small  difference between these simulations and our theoretical predictions can be blamed on the limited dynamic range of the simulations, i.e. $k_cV_A$ not much greater than $\Omega$.  The difference between the reported scalings with resolution seen by Sano et al. and by Fromang and Papaloizou, and the scaling with the gas pressure reported by Sano et al. can be explained as artifacts, due to the numerical methods used in the simulations.

This model also makes specific predictions about numerical simulations of the magnetorotational instability.  First, the magnetic helicity flux carried by small scale structures should arise predominantly from eddies with vertical wavenumbers close to the typical radial wavenumbers.   The strength of this flux should be inversely proportional to $k_c$, the vertical wavenumber cutoff. Second,  vertical turbulent diffusion should arise from slow motions on the scale of the box and should be insensitive to the vertical wavenumber cutoff.  The radial scale of these motions should be close to the box height.  This seems at least qualitatively similar to the numerical results of \citet{JY08}.  The azimuthal scale will be larger by a factor of $\alpha_{ss}^{-1}$, i.e. usually these motions will be axisymmetric due to the effects of shearing.

For real accretion disks this work implies that the behavior of the disk dynamo will be determined by whatever process sets the vertical correlation length of MRI turbulence.  The most important generic effect will be magnetic buoyancy, but the presence of a vertical field may also be important, and can lead to an increased effective $\alpha_{ss}$, especially near the disk midplane.  Microphysical processes that affect magnetic buoyancy may also be important.  For example, in disks with a large radiation pressure, and the diffusion of photons from individual eddies, may increase the role of magnetic buoyancy, broadening eddies and increasing the saturation limit of the dynamo, although the same effect will increase the loss rate of magnetic flux from the disk  Finally, viscous transport coefficients that are too small to suppress MRI by a factor of order unity may lead to an increased dynamo growth rate and a larger $\alpha_{ss}$.

An important implication of this work is that homogeneous shearing box simulations cannot be used to explore the behavior of accretion disks.  Without stratification and magnetic buoyancy the eddy thickness will be determined by numerical effects and the dynamo will be increasingly weak as the numerical resolution is increased.  This same criticism can be applied to claims regarding the role of the magnetic Prandtl number, albeit with less certainty.

Finally, our results suggest a deep link between the MRI driven dynamo and the dynamo driven by the Parker instability.  The scaling arguments presented here do not allow us to conclude which one accounts for the bulk of the magnetic field generation in accretion disks, but they do show that the two dynamos are complementary.  The MRI driven dynamo operates at large values of $L_P$, i.e. near the disk midplane, but with increasing efficiency as we move away from the midplane.  Eventually we come to a region where the MRI and the Parker instability are not cleanly separable from one another, and at larger heights, to a region where the Parker instability drives the disk dynamo.  In this context, the simulations of \citet{JL08}, which show a strong Parker instability driven dynamo in a magnetized accretion disk, are particularly interesting.

\acknowledgments
I am happy to acknowledge the hospitality of the Kavli Institute for Theoretical Physics at UC Santa Barbara where part of this work was done.  I have benefitted greatly from conversations with Axel Brandenberg, Fausto Cattaneo, Julian Krolik, Yoram Lithwick,  Kandaswamy Subramanian, and James Wadsley.  This work has been supported by the National Science and Engineering Research Council of Canada.

\end{document}